\documentclass[aps,prd,letterpaper,onecolumn,amssymb,eqsecnum,nofootinbib]{revtex4}

\usepackage{graphicx}

\usepackage{amsmath}
\usepackage{varioref}
\usepackage{amssymb}
\usepackage[normalem]{ulem}

\begin{document}

\title{Iteration stability for simple Newtonian stellar systems}

\author{Richard H. Price}
\email{rprice@phys.utb.edu}
\affiliation{Center for Gravitational Wave Astronomy, 
80 Fort Brown, Brownsville, TX 78520}

\author{Charalampos Markakis}
\email{markakis@uwm.edu}
\affiliation{
Department of Physics, University of Wisconsin-Milwaukee, P.O.~Box 413,
Milwaukee, WI 53202
}

\author{John L.~Friedman}
\email{friedman@uwm.edu}
\affiliation{
Department of Physics, University of Wisconsin-Milwaukee, P.O.~Box 413,
Milwaukee, WI 53202}

\begin{abstract}

  For an equation of state in which pressure is a function only of
  density, the analysis of Newtonian stellar structure is simple in
  principle if the system is axisymmetric, or consists of a corotating
  binary. It is then required only to solve two equations: one stating
  that the ``injection energy,'' $\kappa$, a potential, is constant
  throughout the stellar fluid, and the other being the integral over
  the stellar fluid to give the gravitational potential.  An iterative
  solution of these equations generally diverges if $\kappa$ is held
  fixed, but converges with other choices. 
  We investigate the
  mathematical reason for this convergence/divergence by starting the
  iteration from an approximation that is perturbatively different 
  from the actual solution. A cycle of iteration
  is then treated as a linear ``updating'' operator, and the
  properties of the linear operator, especially its spectrum,
  determine the convergence properties. For simplicity, we confine
  ourselves to spherically symmetric models in which we analyze
  updating operators both in the finite dimensional space
  corresponding to a finite difference representation of the problem,
  and in the continuum, and we find that the fixed-$\kappa$ operator
  is self-adjoint and generally has an eigenvalue greater than unity;
  in the particularly important case of a polytropic equation of state
  with index greater than unity, we prove that there must be such an
  eigenvalue.  For fixed central density, on the other hand, we find
  that the updating operator has only a single eigenvector, with zero
  eigenvalue, and is nilpotent in finite dimension, thereby giving a
  convergent solution.
\end{abstract}

\maketitle

\section{Introduction}\label{sec:intro}

\subsection{Background}

For a star, or a binary pair of stars, modeled as a barotropic fluid rotating with a known
dependence of the rotation rate $\Omega$ on the radial distance
$\varpi$ from the rotation axis, there are two Newtonian forces per
unit mass acting on a fluid element. The pressure gradient
contributes $-\rho^{-1}\nabla p$ where $\rho$ is the stellar fluid
mass density; gravity contributes $-\nabla \Phi$, where $\Phi$ is the Newtonian
potential. The equivalence of these forces per unit mass to the
centripetal acceleration is
\begin{equation}\label{eq:hydroeq} 
\nabla p+\rho\nabla\Phi=\rho\varpi\,\Omega^2(\varpi)\,.
\end{equation}

For ``cold'' stellar fluid, as in a neutron star, thermal transport is unimportant
and the pressure can be considered to be a function only of the density. The 
force equation (\ref{eq:hydroeq}) can then be written as 
\begin{equation}
  \nabla\left(
h+\Phi-\frac{1}{2}\varpi^2\Omega^2(\varpi)\right)=0\,,
\end{equation}
where $h$, the specific enthalpy, is defined by
\begin{equation}
h(\rho):=\int_0^\rho \frac{dp}{\rho}\,.
\end{equation}
The equations of stellar
structure can then be taken as
\begin{equation}\label{geneq1}
  h+\Phi-\frac{1}{2}\varpi^2\Omega^2(\varpi)=\kappa
  \end{equation}
  \begin{equation}\label{geneq2}
\Phi(\vec{x})=-G\int \frac{\rho(\vec{x}')}{|\vec{x}-\vec{x'}
|}\,d^3x'\,,
\end{equation}
in which the second equation is the Poisson integral for the gravitational 
potential $\Phi$. With $h$ a specified function of $\rho$, these 
two equations constitute the basis for finding $\Phi(\vec{x})$ and $\rho(\vec{x})$,
and hence for solving the problem of stellar structure.

Two essentially different iterative
methods have been used to solve this nonlinear system; each was developed first in
the Newtonian context and then extended to full general relativity: A
Newton-Raphson method was first used by James \cite{james64} and
Stoeckly \cite{stoeckly65} (see also
\cite{eriguchimuller1,eriguchimuller2}). In this method the two
equations (\ref{geneq1}), (\ref{geneq2}) must be solved
simultaneously, requiring the inversion of a large matrix.  The second
method, the so-called self-consistent field (SCF) method, first
introduced in the context of rotating stars by Ostriker and Mark
\cite{ostrikermark}, is much simpler computationally. In this method
one first solves one of the pair (\ref{geneq1}), (\ref{geneq2}) and
then the other.

 In the SCF (as well as the Newton-Raphson) method, iteration requires
fixing two parameters
to be held constant
during the
iteration. Typically one of these is the rotation law, such as the
condition of uniform rotation $\Omega=$\,constant.
The most obvious choice for a second condition would be a fixed value
of the injection energy $\kappa$ in Eq. \eqref{geneq1}. For this
choice, the iteration starts with a guess for a density profile
$\rho(\vec{x})$; Eq. \eqref{geneq2} can then be solved for
$\Phi(\vec{x})$; this result for $\Phi(\vec{x})$ is put in
Eq. \eqref{geneq1} which is solved for $h(\vec{x})$; finally, from the
form of the functional relationship between $\rho$ and $h$, a new,
iterated, solution is found for $\rho(\vec{x})$. This cycle is then
repeated.

It is found that with $\kappa$ held fixed, the SCF iteration does not
converge. For other choices, however, the iteration does converge. For
rotating stars, for example, the SCF iteration converges if the
central density is held fixed. To achieve convergence, Ostriker and
Mark (and many subsequent authors) fix the total mass, while Hachisu
\cite{hachisu86} fixes the ratio of polar to equatorial radius.

Although computational astrophysicists have found success using the
SCF method, no explanation has ever been given for the relationship between
convergence of iteration and
the fixed conditions of the iteration.  Ostriker and
Marck \cite{ostrikermark} mention unpublished work showing stability
for a particularly simple case (the $n=1$ polytrope to be discussed
below) but we are not aware of other analytic studies of iterative
stability of the SCF method of constructing stellar models.

In this paper we provide a mathematical explanation for 
the convergence/divergence properties of the SCF iteration. 
Part of the motivation for doing this is the hope that improved
understanding of the SCF method will lead to simpler 
schemes for
computing the structure of rapidly rotating neutron
stellar models in general relativity.  But most of the motivation
is  curiosity about a feature of computations that has been
known but unexplained for more than forty years.

We have approached this problem by considering the iteration 
to start very close to a solution of Eqs.~\eqref{geneq1},~\eqref{geneq2}. That is,
we start with an initial guess for $\rho(\vec{x})$ that is 
perturbatively different from the exact function that would
solve Eqs.~\eqref{geneq1},~\eqref{geneq2}. The exact structure problem ~\eqref{geneq1},~\eqref{geneq2}
is then replaced by the equations for  $\delta h$,
$\delta \Phi$, $\delta\kappa$, $\delta\rho$,
the differences between the true values of $h,\Phi,\kappa,\rho$
and the values at the start of an iteration cycle. Higher order
terms in these perturbations are dropped, resulting 
in linear equations for the perturbations 
\begin{equation}\label{itpairpert}
  \delta h+\delta\Phi=\delta\kappa
\quad \quad\quad
\delta \Phi(\vec{x})=-G\int \frac{\delta \rho(\vec{x}')}{|\vec{x}-\vec{x'}
|}\,d^3x',\,
\end{equation}
with the rotation law fixed during the iteration, $\delta\Omega=0 $. A cycle of iteration then starts with an initial perturbation $\delta \Phi$, and
ends with an iterated perturbation $\delta \Phi$.  One round of
iteration, therefore, constitutes a linear ``updating'' of the density
perturbation $\delta\rho(\vec{x})$. Whether or not this linearized
iteration converges depends on the properties of the linear operator 
that performs the updating.

It is fairly clear that convergence of the linearized problem of
Eq.~(\ref{itpairpert}) is a necessary condition for convergence of the
SCF iterative solution to the exact equations
\eqref{geneq1},~\eqref{geneq2}.  Our working
assumption is that it is also a sufficient condition; i.e.\,, if the
linearized iteration converges, then iteration of the exact problem
will also converge if the iteration is started close enough to the
exact solution.  Our numerical results support this assumption: exact
iteration converges if and only if linearized iteration converged.

The analysis of convergence then becomes a study of the properties of
the linear updating operator. We find that for fixed-$\kappa$
iteration the linear updating operator is self-adjoint and (not
surprisingly) generally has an eigenvalue larger than unity. Less
expected is the result of the study of the cases in which iteration
does converge. Here it is found that the eigenbasis is not complete,
but (for a finite difference representation of the equations) is
nilpotent, and can usefully be understood with a Jordan decomposition
into generalized eigenvectors.

The evidence for these explanations, for both rotating stars and
binaries, is extensive, but consists largely of numerical results.
The simplicity of the spherical case, on the other hand, allows some
very definitive and interesting mathematical results and physical
insights, and it is that case that we present here.

\subsection{Spherically symmetric equations}

In spherical symmetry the structure problem is that of determining $\Phi(r)$ 
and $\rho(r)$, where $r$ is the radial distance from the center of symmetry.
Since $\Omega=0$ in the spherical case, the equation of hydrodynamic
equilibrium reduces to 
\begin{equation}
  \label{eq:spheq1}
  h+\Phi=\kappa\,.
\end{equation}
It is convenient to consider the enthalpy $h$, rather than 
the density, to be the unknown structure function, so we assume that $\rho$ is
a known invertible functional of $h$ given by 
\begin{equation}
  \label{eq:calHdef}
\rho=\varrho[h]\,.
\end{equation}
In the particular, and very common case of a polytropic equation of state,
\begin{equation}\label{eq:polyEOS} 
p=K\rho^{1+1/n}\,,
\end{equation}
the enthalpy function is
\begin{equation}\label{eq:polyh} 
h=K(1+n)\rho^{1/n}\,,
\end{equation}
so that 
\begin{equation}\label{eq:polyEOS2}
  \varrho(h)=\left[h/K(1+n)\right]^n.
\end{equation}

The Poisson equation for spherical symmetry, in terms of $h$, is
\begin{equation}
  \label{eq:structc}
\Phi(r)=-4\pi G\left[\frac{1}{r}\int_0^r\varrho[h(r')]r'^2\,dr'
+\int_r^R\varrho[h(r')]r'\,dr'
\right]\,,
\end{equation}
and we assume that $\rho$ is finite at $r=0$, so that $\Phi(0)$
is finite.
With differentiation this system can be cast as the following 
differential equation for $h(r)$:
\begin{equation}
  \label{eq:diffeq}
\frac{1}{4\pi G r^2}  \frac{d}{dr}\left(
r^2\frac{dh}{dr}
\right)=-\varrho[h(r)]\,.
\end{equation}
Equation (\ref{eq:structc}) is equivalent to this differential equation
with the constraint that $h(r)$ be analytic at $r=0$.
The value $r=R$ at which $h$ first vanishes
determines the radius $R$ of the stellar surface and the boundary of the
domain in which Eq.~(\ref{eq:spheq1}) applies.

The linearization of the spherical problem leads to the 
equations
\begin{equation}
  \label{eq:gendeltareln}
  \delta h+\delta \Phi=\delta\kappa
\end{equation}
\begin{equation}
  \label{eq:gendeltaU}
\delta \Phi
=-4\pi G\left[\frac{1}{r}\int_0^r {\cal P}(r')\delta h(r')^{\rm old}\,
r'^2\,dr'+\int_r^R{\cal P}(r')\delta h(r')^{\rm old}\,
r'\,dr'
\right]
\,,
\end{equation}
in which 
\begin{equation}\label{Hdef}
  {\cal P}:=d \varrho/dh.
\end{equation}In Eq.~(\ref{eq:gendeltaU}) we have have ignored changes in the
stellar radius $R$, because their contribution can be shown to be of order higher than linear, as long as  the star is compressible\footnote{
For a polytrope of index $n$, the  contribution in Eq. \eqref{eq:gendeltaU} of the change $\delta R$ in radius can be shown to be of order $1+n$ in $\delta R$.}.
 The basis of our analysis is  Eqs.~(\ref{eq:gendeltareln}) and (\ref{eq:gendeltaU}), with different
choices of the form of ${\cal P}$, and different choices of what is
held fixed during iteration.

 \subsection{Outline}

Our study of the SCF solution of spherical structure was guided by,
and aimed at explaining the known numerical phenomena in nonspherical
SCF iteration, along with our own numerical discoveries. These come
almost entirely from computations on a finite difference grid with
barotropic equations of state and include the
following. (i)~Iteration with fixed $\kappa$  appears always to
diverge. (ii)~For rotating stellar models, iteration with fixed
central density always converges.

For spherical stellar models, by analyzing the {\em mathematical}
properties of the linearized updating operator (as opposed to studying
its numerical results), we have been able to show the following:
(i)~For iteration with $\kappa$ fixed, the linearized updating operator
is self-adjoint and there is a complete basis of density
perturbations. (This set of perturbations is {\em not} related to the
likewise complete and orthogonal set of density perturbations that are
eigenmodes of radial oscillations of spherical stellar models.)  For
models with polytropic index $n$ we have shown that there is a single
eigenvector with eigenvalue greater than $n$, and that all others must
have eigenvalues less than $n$.  This discovery motivated a study of
models with $n<1$, and it was found that, indeed, for sufficiently
small (and astrophysically irrelevant) $n$, the fixed-$\kappa$
operator {\em does} have all linearized eigenvalues less than unity,
and iteration does converge.  (ii)~For fixed central density on a
finite difference grid, there is only a single eigenvector and the
single eigenvalue zero. This spectrum has a very clear physical
explanation. For the continuum version of the linearized updating
operator for fixed central density, we show that iteration must
converge, and that there can be no bounded eigenvector. The single
eigenvector of the discrete implementation corresponds, in the continuum, to a delta
function. (iii)~For iteration with fixed density
 at some radius $r_{f}>0$ that is less than the stellar radius $R$, we
show that there is an infinite number of eigenvectors, that these
eigenvectors do not form a complete basis, and that the updating may or 
may not converge.  In the finite difference implementation of this  problem, with $N$ grid zones in the interval $(0,R)$, the number of eigenvectors is equal to the number of grid zones in the interval $(0,r_{f})$.

The remainder of this paper is organized as follows.  In
Sec.~\ref{sec:constkappa} we present the specialization of the
spherically symmetric iteration scheme to the fixed-$\kappa $
case. We show that the linearized updating operator is self adjoint
(for a suitably chosen function space and inner product). We derive a
strict bound on the eigenvalues in the case of a polytropic equation
of state, and (less useful) bounds for more general equations of
state.  We turn, in Sec.~\ref{sec:constrhoc}, to iteration with
density fixed at some radius $r_{f}$.  Finite difference results are
given that show for $r_{f}=0$ that there is only a single eigenvector,
with zero eigenvalue. For $0<r_{f}<R$, we show that the number of
eigenvectors is proportional to the choice of $r_{f}$. We give a
physical explanation of these numerical results and then consider the
equivalent problem in the continuum.  We show that for $r_{f}=0$ the
iteration must converge, and we show that there can be no bounded
eigenvector. We go on to show that for $r_{f}\neq0$ the iteration
problem has aspects both of the fixed-$\kappa$ problem
 and of the fixed central
density problem.  The paper is summarized, and conclusions given, in
Sec.~\ref{sec:summary}.

\section{Fixed-$\kappa$ iteration}
\label{sec:constkappa}

\subsection{Self-adjoint linearized updating operator}

If we set $\delta\kappa=0$ in Eqs.~(\ref{eq:gendeltareln}), (\ref{eq:gendeltaU})
then $\delta h=-\delta \Phi$ and we have
\begin{equation}
  \label{eq:fixkapA}
\delta h^{\rm new}(r)
=4\pi G\left[\frac{1}{r}\int_0^r {\cal P}(r')\delta h(r')^{\rm old}\,
r'^2\,dr'+\int_r^R{\cal P}(r')\delta h(r')^{\rm old}\,
r'\,dr'
\right]\equiv L_{\kappa}(\delta h^{\rm old})
\,,
\end{equation}
where $L_\kappa$ is the linear updating operator for fixed
$\kappa$.  The inverse of this linearized updating operator is found,
by differentiating Eq.~(\ref{eq:fixkapA}) , to be
\begin{equation}
  \label{eq:inverseL}
L_\kappa^{-1}\big(v(r)\big) = -\frac{1}{4\pi Gr^2 {\cal P}(r)}\,\frac{d}{dr}\left(
r^2\frac{dv}{dr}
\right)\,.
\end{equation}
Any smooth function $v(r)$ that results from an application of $L_\kappa$ in 
Eq.~(\ref{eq:fixkapA}) will have the property  at the stellar 
surface $r=R$ that
\begin{equation}
  \label{eq:surfcond}
\left.\frac{dv}{dr}\right|_{r=R^-} =- \left.\frac{v}{r}\right|_{r=R^-}\,.
\end{equation}
We take this to be one of the conditions on the function space on 
which $L_\kappa$ and $L_\kappa^{-1}$ operate. The second condition is that $v(r)$
must be finite at the stellar center $r=0$.

As  our inner product on this space we choose
\begin{equation}
  \label{eq:dotdef}
  v_1\cdot v_2=\int_0^R r^2{\cal P}(r)v_1(r) v_2(r)\,dr\,.
\end{equation}
from which we get
\begin{equation}
  \label{eq:boundary}
  v_1\cdot L_\kappa^{-1}(v_2)  -v_2\cdot L_\kappa^{-1}(v_1)  =- \frac{1}{4\pi G}
  \left[
    v_1r^2\frac{dv_2}{dr}-v_2r^2\frac{dv_1}{dr}
  \right]_0^R\,.
\end{equation}
With the conditions that the functions $v_1,v_2$ are well behaved at
$r=0$ and satisfy Eq.~(\ref{eq:surfcond}), the right hand side above
vanishes and we conclude that $L_\kappa^{-1}$ is self-adjoint\footnote{
We use ``self-adjoint'' in the physicist's sense of an
operator symmetric with respect to the inner product defined above,
 when the operator is restricted to smooth functions.  We do not
 characterize its domain.
}. We
shall see that $L_\kappa^{-1}$ has no zero eigenvalues, and hence is
invertible, and $L_\kappa$, its inverse, is self adjoint. (We note here
that for nonspherical models it is equally simple to prove that the fixed-$\kappa$ updating operator is self-adjoint.)

\subsection{Polytropes and eigenvalue bounds}\label{sec:polybound}

We now specialize to the polytropic equations of state. In this case, 
the $\varrho$ function in 
Eq.~(\ref{eq:polyEOS2})
is used in Eq.~(\ref{eq:diffeq}).
To simplify notation we follow common convention\cite{Chandrasekhar1939} 
and introduce dimensionless variables
\begin{equation}
  \label{eq:dimnless}
  \Theta\equiv\frac{h}{h_0}\quad\quad\quad \xi\equiv r\sqrt{
\frac{
4\pi G}{h_0} \varrho(h_0)\;}
\,= r\sqrt{
\frac{
4\pi G}{h_0} \;}
\,\left(
\frac{h_0}{K(1+n)}
\right)^{n/2}
\,,
\end{equation}
in which $h_0$ is the unperturbed value of $h$ at the stellar center $r=0$.  In
terms of these variables the nonlinear equation of structure
Eq.~(\ref{eq:diffeq}) becomes the Lane-Emden equation\cite{Chandrasekhar1939}
\begin{equation}
  \label{eq:laneembden}
  \frac{1}{\xi^2}\frac{d}{d\xi}\left( \xi^2\frac{d\Theta}{d\xi} \right)
=-\Theta^n\ .
\end{equation}

For fixed-$\kappa$ perturbations about a solution of
Eq.~(\ref{eq:structc}), the eigenvalue problem $L_\kappa(v)=\lambda v$
for Eq.~(\ref{eq:fixkapA}) is equivalent to solving the inverse
problem $L_\kappa^{-1}(v)=(1/\lambda) v$ for the differential operator
$L_\kappa^{-1}$ in Eq.~(\ref{eq:inverseL}). With ${\cal P}\equiv
d\varrho/dh=nh^{n-1}/[K(1+n)]^n$, and with the dimensionless radial
variable $\xi$ this becomes
\begin{equation}
  \label{eq:pertA}
  \frac{1}{\xi^2}\frac{d}{d\xi}\left( \xi^2\frac{dv}{d\xi} \right)
=-\frac{n}{\lambda}\left(\frac{h}{h_0}\right)^{n-1}  v\,.
\end{equation}

We next introduce the notation $f\equiv\xi\Theta$ for the Lane-Emden
equation, and $F\equiv\xi v$ for the perturbation equation, and we
rewrite these equations, respectively, as
\begin{eqnarray}
  \label{eq:modeqA}
\frac{d^2f}{d\xi^2}&=&-\left(\frac{f}{\xi}\right)^{n-1}   f\\
  \label{eq:modeqB}
  \frac{d^2F}{d\xi^2}&=&-\frac{n}{\lambda}\left(\frac{f}{\xi}\right)^{n-1} F\ .
\end{eqnarray}
In the first of these, $f(\xi)$ is to be considered an unknown
function to be found on $0<\xi<\xi_{\rm max}$ by solving the equation
subject to the normalization  $f=\xi+{\cal O}(\xi^3)$. In the
second equation $f/\xi$ is considered to be a known function of $\xi$,
and the eigenequation is to be solved subject to the normalization
$F=\xi+{\cal O}(\xi^3)$ and
\begin{equation}
  \label{eq:Fbc}
\left.  dF/d\xi \right|_{\xi=\xi_{\rm max}}=0\,,
\end{equation}
which is equivalent to Eq.~(\ref{eq:surfcond}).

\begin{figure}[ht]
\centering
\includegraphics[width=.6\textwidth]{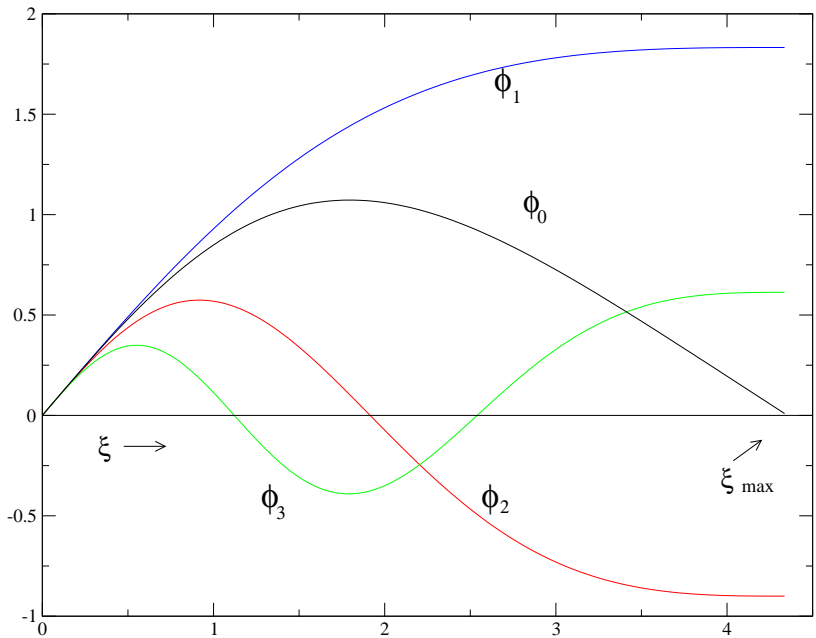}
\caption{Various solutions of 
$\phi_{,\xi\xi}=-G(\xi)\,\phi$ for $n=2$.
}
\label{fig:curves}
\end{figure}

With a solution for $(f/\xi)^{n-1}\equiv\Theta^{n-1}\equiv G(\xi)$ treated as a known 
function, we can view both
Eqs.~(\ref{eq:modeqA}) and (\ref{eq:modeqB}) as particular cases of the equation
\begin{equation}
  \label{eq:modeqC}
  \frac{d^2\phi}{d\xi^2}=-k G(\xi)\,\phi\,,
\end{equation}
with normalization $\phi=\xi+{\cal O}(\xi^3)$\,.  For this
equation we know that if $k=1$, then $\phi$ vanishes at $\xi=\xi_{\rm
  max}$ and is positive for $\xi<\xi_{\rm max}$.  In
Fig.~\ref{fig:curves}, solutions of Eq.~(\ref{eq:modeqC}) are given in
which $G(\xi)$ corresponds to the particular case $n=2$.  The curve
$\phi_0$ shows the solution for $k=1$, i.e., the solution
corresponding to the equilibrium profile of the star.  The figure also
shows eigenfunctions, solutions corresponding to condition
(\ref{eq:Fbc}). Shown are the eigenfunction $\phi_1$ for the lowest
eigenvalue, $\phi_2$ for the next higher eigenvalue, and $\phi_3$ for
the next.  It seems intuitively clear that the eigenvalue $k$ for
$\phi_1$ must be less than unity, since Eqs.~(\ref{eq:Fbc}) and
(\ref{eq:modeqC}) require that $\phi_1$ be ``less curved'' than
$\phi_0$. Similarly, the eigenvalue for $\phi_2$, and all other
eigenfunctions, must be larger than unity.  We now prove that this
must be so.

We start by proving that the zeroes and extrema of a smooth solution
to Eq.~(\ref{eq:modeqC}) must alternate. Consider two extrema of a
solution.  There must be a point between those two extrema at which
$d^2\phi/d\xi^2=0$. But Eq.~(\ref{eq:modeqC}) requires that $\phi=0$
at that point. Between any two zeros, of course, there must be an
extremum. Thus zeroes and extrema alternate, as claimed.
We next consider two functions $\phi_A$ and $\phi_B$ that are
solutions of Eq.~(\ref{eq:modeqC}), with the starting condition
$\phi=\xi+{\cal O}(\xi^3)$. Let these two solutions (not necessarily
eigenfunctions) correspond respectively to $k_A$ and $k_B$, with
$k_A>k_B$. Suppose that there is some point $\xi=a$ such that
$\phi_A\geq\phi_B\geq0$ for $0<\xi\leq a$. We then consider
\begin{equation}
  \label{eq:lemma}
\left(  \phi_B-\phi_A\right)_{,\xi\xi}=k_A\phi_A-k_B\phi_B\quad\quad
\mbox{for $0<\xi\leq a$}\ .
\end{equation}
By our assumptions, the right hand side is everywhere positive.  But
the function $\phi_B-\phi_A$ starts with value zero at $\xi=0$ and with
a zero derivative. It follows from Eq.~(\ref{eq:lemma}) that
$\phi_B-\phi_A$ must be positive for $\xi\leq a$, which contradicts our
assumptions that $\phi_A\geq\phi_B\geq0$ for $0<\xi\leq a$.
 This proves that there can be no interval $0<\xi\leq a$ on
which $\phi_A>\phi_B>0$. Essentially the same argument shows that $\phi_B$ is
positive at the first zero of $\phi_A$, and more generally that the
$\xi$ value of the
first zero of a solution of Eq.~(\ref{eq:modeqC}) (subject to the
boundary conditions at $\xi=0$) decreases as $k$ increases. This
immediately confirms that an eigenfunction, like $\phi_1$, with no
zero, must correspond to a value of $k$ less than unity.
We also conclude that $\phi_2$ and $\phi_3$, and any eigenfunction that has
extrema intermediate between 0 and $\xi_{\rm max}$, must have a zero
for $\xi<\xi_{\rm max}$, and hence an eigenvalue $k$ that is larger
than unity.  This completes the proof that there is one and only one
eigenvalue $k$ that is smaller than unity.

When applied to the problem of Eqs.~(\ref{eq:laneembden}) and
(\ref{eq:pertA}), this tells us that there is one and only one
eigenvalue $\lambda$ of the updating operator that is larger than the
polytropic index $n$.  For astrophysically relevant models, which have
$n>1$, this guarantees that fixed-$\kappa$ iteration will have an
eigenvalue of the updating operator that is larger than unity, and
that updating will not converge. It suggests, but does not guarantee
that there will only be a single updating eigenvalue $\lambda$ that is
larger than unity. This, however, does turn out to be what we have found
in numerical studies (see below).

Though a polytropic equation of state with $n<1$ is not
astrophysically plausible, it is interesting since the above analysis
shows that fixed-$\kappa$ iteration need not diverge for $n<1$. We
have found numerically that for $0<n<0.016$ the nonlinear fixed-$\kappa$
iteration is, indeed, convergent.


A particularly simple example of the above analysis is
the case $n=1$, for which the Lane-Emden equation (\ref{eq:laneembden}) admits an
analytical solution
\begin{equation}  \label{eq:theta1}  
\Theta(\xi)=\frac{\sin\xi}{\xi}
\end{equation}
which vanishes at the surface $\xi_{\max}=\pi$.
In this case the eigenvalue problem of Eq.~(\ref{eq:pertA})
is a spherical Bessel differential equation
\begin{equation}
  \label{eq:Bessel0}
-\frac{1}{\xi^2}\frac{d}{d\xi}\left( \xi^2\frac{dv}{d\xi} \right)
=\frac{1}{\lambda} v\,,
\end{equation}
and is analytically tractable.
 The
solutions regular at the origin are
zeroth order spherical Bessel functions:
\begin{equation}  \label{eq:eigenfunctions1}  
v^{(k)}(\xi)=a_{k}\,\ \frac{\sin(\lambda_{k}^{-1/2}\xi)}{\xi}\quad 
\end{equation}
where $k$ numbers the eigenfunctions and the $a_{k}$ are normalization
constants. To satisfy the  boundary condition (\ref{eq:surfcond}), the eigenvalues are required to have the values
\begin{equation}  \label{eq:eigenvalues1}  
\lambda_{k}=\left(k-\tfrac{1}{2}\right)^{-2},\quad k=1,2,\ldots .
\end{equation}
 
\begin{figure}[ht]
\centering
\includegraphics[width=.6\textwidth]{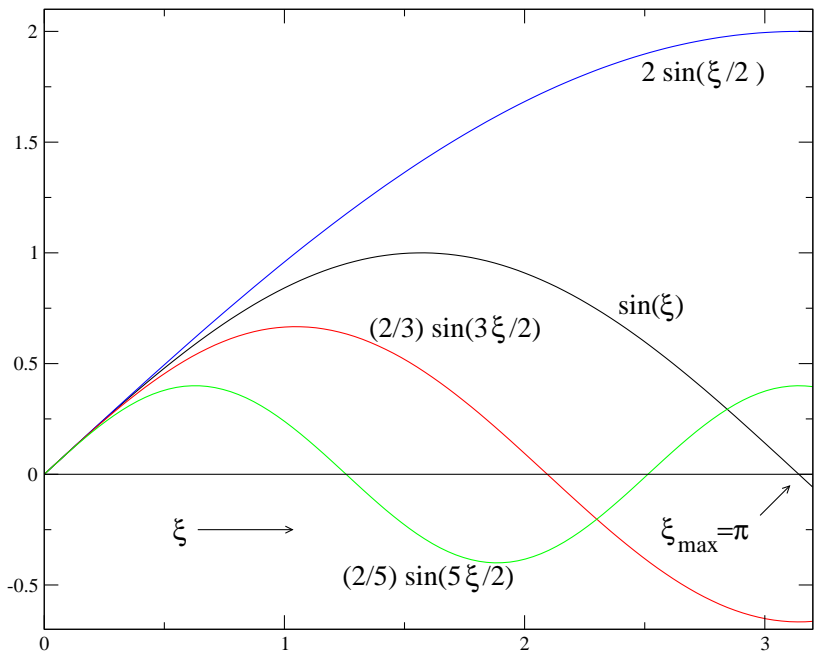}
\caption{For $n=1$  the linearized problem 
has  simple closed form solutions to the problem illustrated in 
Fig.~\ref{fig:curves}. In this case the solution of the Lane-Emden
equation gives $\xi_{\rm max}=\pi$ and $\phi_0=\sin\xi$. With the eigensolutions
$\phi_k$
scaled to have
$\phi=\xi+{\cal O}(\xi^3)$
the first three eigenmodes are $\phi_1=2\sin{(\xi/2)}$;
$\phi_2=(2/3)\sin{(3\xi/2)}$; and $\phi_3=(2/5)\sin{(5\xi/2)}$.}
\label{fig:v123}
\end{figure}

The resulting functions $\phi_k(\xi):=\xi v^{(k)}(\xi)$ are plotted in
Fig.~\ref{fig:v123}.  It is simple to check that the eigenfunctions
satisfy the orthogonality
condition\begin{equation} \label{eq:innerorthonormal} v^{(k)}\cdot
  v^{(k')} \equiv \int_0^\pi\ \xi^2v^{(k)}(\xi) v^{(k')}(\xi)\,d\xi\ =
  0 \quad\mbox{if $k\neq k'$} \,.
\end{equation}

For a nonpolytropic equation of state, we can arrive at somewhat weaker
results for bounds on the eigenvalues. From Eq.~(\ref{eq:diffeq})
and the eigenproblem associated with its linearization we have 
\begin{eqnarray}
  \frac{d^2f}{dr^2}&=&-A\,f\label{nonpoly}\\
\frac{d^2F}{dr^2}&=&-\frac{1}{\lambda} B F\,.\label{nonpolyeigen}
\end{eqnarray}
Here $f\equiv rh$, $F\equiv r\delta h$ and 
\begin{equation}
  \label{eq:AandBdefs}
A\equiv4\pi G\varrho/h\quad\quad\quad B\equiv 4\pi G d\varrho/dh\,.   
\end{equation}
Equation (\ref{nonpoly}) is solved and $r_{\rm max}$ is defined as the
first zero of the solution. That solution is then used in $B$, so that
Eq.~(\ref{nonpolyeigen}) is considered to be a linear eigenproblem for
$F$.  The boundary condition on that equation is $dF/dr=0$ at
$r=r_{\rm max}$. 

With only slight modification, the argument used in the polytropic
case can be used to show: (i)~If there is a constant $c_1$ such that
$B>c_1A$ for all $r\in(0,r_{\rm max})$, then the largest $\lambda$ must 
be greater than $c_1$. (ii)~If there is a constant $c_2$ such that 
$B<c_2A$ for all $r\in(0,r_{\rm max})$, then all $\lambda$s except 
the largest, must be less than $c_2$.

\subsection{Numerical investigations}\label{sec:fixkappaFD}

In terms of the polytropic variables of Eq.~(\ref{eq:dimnless}), the
updating equation (\ref{eq:fixkapA}) can be compactly written in the
dimensionless form,
\begin{equation}
  \label{eq:hupdatedimnles}
  \delta h^{\rm new}({\xi})= n\int_0^{\xi_{\max}}\ 
  \Theta(\xi')^{n-1}\frac{\delta h(\xi')^{\rm old}\,}{\rm max(\xi,\xi')}
  \,\xi '^2\,d\xi'\equiv L_{\kappa}(\delta h^{\rm old})\,,
\end{equation}
which motivates the eigenvalue problem
\begin{equation} \label{eq:veigenk}
L_{\kappa}(v)\,\equiv n\int_0^{\xi_{\max}}\ \Theta(\xi')^{n-1}\frac{ v(\xi')
}{\rm max(\xi,\xi')}\,\xi '^2\,d\xi' =\lambda v(\xi)\,.
\end{equation}
This integral form is equivalent to inversion of the differential
equation (\ref{eq:pertA}) under the boundary conditions
(\ref{eq:surfcond}). The stability of the linearized iteration
(\ref{eq:hupdatedimnles}) depends on the spectrum of the linear
operator $L_{\kappa}$. The spectrum is obtained by solving
Eq.~(\ref{eq:veigenk}), which  requires that the background
(unperturbed) solution $\Theta(\xi)$ be known.

Analytical solutions to the Lane-Emden equation
\cite{Chandrasekhar1939} exist only for polytropic indexes $n=0, 1,
5$; for other values of $n$, we have solved
Eq.~(\ref{eq:laneembden}) numerically for $\Theta(\xi)$, using a
predictor-corrector Adams method of adaptive stepsize and order. The
eigenvalue problem of Eq.~(\ref{eq:veigenk}) is discretized on the
grid of radial coordinates $\xi$ of Eq.~(\ref{eq:dimnless}) using $N$
equidistant grid points in the star interior:
\begin{equation} \label{eq:ksigrid}
  \xi_i=\left(i-\tfrac{1}{2}\right)\Delta\xi,\quad\quad i=1,...,N, 
\end{equation} 
where
\begin{equation}  \label{eq:ksispacing}  
\Delta\xi=\xi_{\max}/N
\end{equation}
is the grid spacing. Upon discretization, Eq.~(\ref{eq:veigenk})
becomes:
\begin{equation}
  \label{eq:Lijvj}  
\sum _{j=1}^N (L_\kappa)_{ij}v_{j}=\lambda\\v_{i} 
\end{equation}
where $v_i\equiv v(\xi_i)$ are the components of a vector
${\mathbf{v}}$ formed from the values of the eigenfunction at
the grid points, and
\begin{equation} \label{eq:Lij} (L_\kappa)_{ij}=n
  \Theta(\xi_j)^{n-1}\frac{\xi_{j}^2}{{\max}(\xi_{i},\xi_{j})}\,\Delta\xi\,.
\end{equation}The matrix $S_{ij}\equiv1/{{\max}(\xi_{i},\xi_{j})}$ is symmetric,
which means that $(L_\kappa)_{ij}$ has the form
\begin{equation}
  \label{eq:symtimesdiag}
  (L_\kappa)_{ij}=\sum _{k=1}^N S_{ik}D_{kj} 
\end{equation}
where $\mathbf{D}$ is a diagonal matrix with positive entries.  Since the
diagonal elements of $\mathbf{D}$ are positive, the matrix $\mathbf{D}^{1/2}$, defined
by $\mathbf{D}^{1/2}\mathbf{D}^{1/2}=\mathbf{D}$, is real and has inverse
$\mathbf{D}^{-1/2}$. The nonsingular similiarity tranformation $\mathbf{D}^{1/2}\mathbf{L_{\kappa}D}^{-1/2}=$
$\mathbf{D}^{1/2}(\mathbf{S D})\mathbf{D}^{-1/2} =\mathbf{D}^{1/2}\mathbf{S D}^{1/2}$ symmetrizes $\mathbf{L}_{\kappa}$, proving that $\mathbf{L}_{\kappa}$
has a complete basis of real eigenvectors.  Iteration will converge if and only 
if all the eigenvalues have magnitude less than unity.

\begin{figure}[ht] \centering
\includegraphics[width=.48\textwidth]{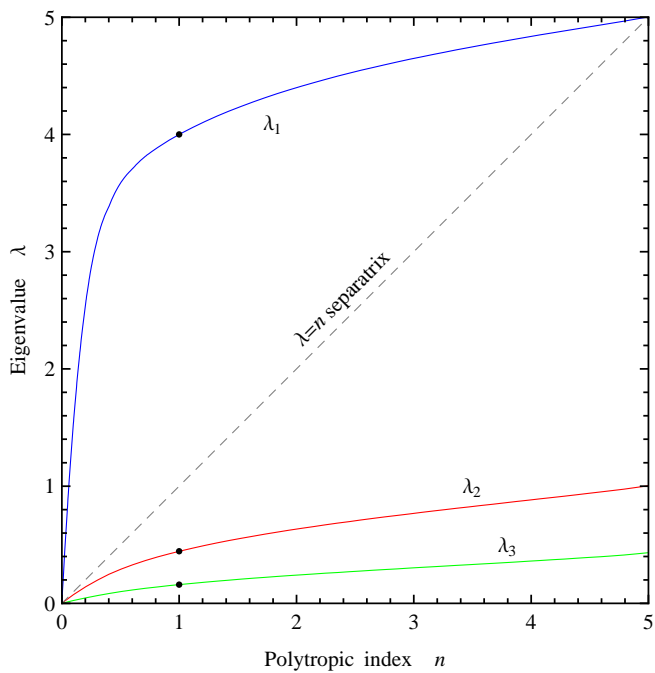}
\caption{Eigenvalues $\lambda$ for various  polytropic indices
  $n$. The maximum eigenvalue $\lambda_{1}$ is separated from
  $\lambda_{2}, \lambda_{3},...$ by the separatrix $\lambda = n$
  (dashed line). Black dots denote the eigenvalues \eqref{eq:eigenvalues1} that correspond to the modes of Fig.~\ref{fig:v123}.}
\label{fig:lamda123}
\end{figure}

Once the background solutions $\Theta(\xi)$ are known from a numerical
solution of Eq.~(\ref{eq:laneembden}), their values are interpolated
to the grid points \eqref{eq:ksigrid}, and the numerical computation
of the eigenvalues of the matrix (\ref{eq:Lij}) for various values of
$n\in(0,5)$ is straightforward. (For $n\geq5$ the value of $\Theta$
never goes to zero; that is,  the stellar model it represents has infinite
radius. In this case, $\xi_{\max}$ does not exist, and the eigenvalue
problem is not defined. For $n=0$ the stellar fluid is incompressible
and the background solution is not smooth at the stellar surface.) For
sufficiently large $N$, these eigenvalues should approach those of the
continuum operator. The results of such a computation are plotted in
Fig.~\ref{fig:lamda123}, and confirm that one and only one eigenvalue
$\lambda$ is greater than $n$, while all others are lower than $n$.

To check a nonpolytropic equation of state we used 
\begin{equation} \label{eq:whitedwarfEOS}
  \rho=\varrho(h)=(ah+bh^2)^{3/2}.
\end{equation}
which approximates white dwarf equations of state
\cite{Chandrasekhar1939}. As one might expect, for $a\gg b h_0$ the
unperturbed solution approaches an $n=3/2$ polytrope, while for $a\ll
b h_0$ the unperturbed solution approaches an $n=3$ polytrope. It is
also known \cite{Chandrasekhar1939} (page 430) that generic solutions
for the equation of state \eqref{eq:whitedwarfEOS}, with arbitrary
values of $bh_{0}/a$, are bounded by the aforementioned
polytropes. Upon computation, we found that the eigenspectrum of
$L_{\kappa}$ also exhibits a similar behavior.  The eigenvalues of
this problem vary monotonically between the $n=3/2$ and the $n=3$
eigenvalues, as $bh_{0}/a$ varies from 0 to $\infty$. One may thus be
able to infer the nature of the eigenspectrum -- and convergence
vs. divergence -- by considering polytropic equations of state that in
some sense ``bound" the actual equation of state.

\section{Fixed  density} 
\label{sec:constrhoc} 

\subsection{Fixed central density and finite difference computations }\label{subsecthreeA}

As a specific case of conditions that lead to convergent iteration, we
consider fixed density at some radius. Here we start with the simplest
case: fixed density at the stellar center.  We have studied
convergence of the SCF\ iteration for spherically symmetric polytropic
models with a wide range of indices. To check a nonpolytropic equation
of state we again used Eq. \eqref{eq:whitedwarfEOS}.  In all cases we
found that the iteration converged. We now analyze why this is so by
considering small deviations from the solution.

For spherical models with fixed central density, $\delta
h=-\delta\Phi+\delta \Phi|_{r=0}$ in Eqs.~(\ref{eq:gendeltareln}),
(\ref{eq:gendeltaU}), and we get
\begin{equation}\label{eq:fixcentral1}
  \delta h^{\rm new}(r)
 =4\pi G\int_0^R {\cal P}(r')\delta h(r')^{\rm old}\left(\frac{1}{\max(r,r')}
-\frac{1}{r'}
\right)
r'^2\,dr'
  \equiv L_{\rm cent}(\delta h^{\rm old})
  \,.
\end{equation}
Although the updating operator $L_{\rm cent}$ is superficially similar to 
$L_\kappa$ in
Eq. \eqref{eq:fixkapA}, 
the operator $L_{\rm cent}$ is not self-adjoint and, as 
we shall now demonstrate, its eigenspectrum
is dramatically different from that of $L_\kappa$.

The properties of this updating operator become particularly
transparent upon discretization. With a discretization that is a
modification of that  in
Sec.~\ref{sec:fixkappaFD}, 
\begin{equation}\label{rdiscrete}
  \Delta r=R/N, \quad\quad r_j=(j-\textstyle{\tfrac{1}{2}})\Delta r, 
\quad\quad j=1,...,N 
\end{equation}
the eigenvalue problem for $L_{\rm cent}$
becomes
\begin{equation} \label{eq:LcentDiscreteEigenproblem}  
\sum_{j=1}^N (L_{\rm{cent}})_{ij}v_{j}=\lambda\\v_{i}
\end{equation}
with
\begin{equation} \label{eq:LcentDiscreteDefinition}
  (L_{\rm{cent}})_{ij}=4\pi G {\cal P}(r_j)\, \left(\frac{1\,}{{\rm
        max}(r_i,r_{j})}\,-\frac{1\,}{r_j}\right)r_{j}^2\Delta r\,. 
\end{equation} 
We note that $(L_{\rm{cent}})_{ij}=0$ for $r_i\leqslant r_{j}$
or, equivalently, $i\leqslant j$.  This means that the matrix  $\mathbf{L}_{\rm{cent}}
\in\mathbb{R}^{N\times N}$ is  {\it strictly} lower triangular, that is, 
a lower triangular matrix with zeroes on the diagonal. It
 follows that
\begin{equation}  \label{eq:LcentDet}  
{\rm{det}}(\mathbf{L}_{\rm{cent}}-\lambda\mathbf{I})=\lambda^{N}=0\,,
\end{equation}
so that the only eigenvalue can be zero.  Below the diagonal, no
element is zero, and each column has a different length of nonzero
entries. (Note that neither $r=0$ nor $r=R$ is included in the grid,
so ${\cal P}(r_j)$ is always nonzero.) It follows that there are $N-1$
linearly independent columns, and hence that the rank of the matrix is
$N-1$, and therefore that there is only a single zero eigenvector.  It
is easily seen that, modulo scaling, this eigenvector is
\begin{equation}  \label{eq:ZeroEigenvector}  
v^{(1)}_j=\delta_{jN}
,\end{equation}
since this satisfies Eq. \eqref{eq:LcentDiscreteEigenproblem},
with vanishing eigenvalue:
\begin{equation} \label{eq:ZeroEigenvectorEquation} \sum_{j=1}^N
  (L_{\rm{cent}})_{ij}v^{(1)}_j=\sum_{j=1}^N
  (L_{\rm{cent}})_{ij}\delta_{jN}=(L_{\rm{cent}})_{iN}=0\,.
\end{equation}

The convergence of the linearized iteration is obvious from the
strictly lower triangular nature of $\left( L_{\rm{cent}}
\right)_{ij}$: When this matrix is applied to any column vector, the
result is a column with leading entry zero. A second application gives
zero for the first two elements of the column, etc. It is obvious that
the matrix is therefore nilpotent of index $N$.  Not only is iteration with
$\left( L_{\rm{cent}} \right)_{ij}$ convergent, it reduces any initial
perturbation to zero after $N$ iterations.  This suggests why the SCF
method of solving may be so successful.

Though a complete eigenbasis does not exist, one can construct a basis
of generalized eigenvectors, by putting the matrix for $L_{\rm{cent}}$
in Jordan canonical form \cite{jordan,strang}, an approach that will be useful
for nonspherical models. 
In this block-diagonal form, the subspace corresponding to each block
contains a single eigenvector. In our spherically symmetric case there
is only a single eigenvector in the whole space, so the Jordan
canonical form consists of a single block. The basis vectors in the
canonical form, ${\mathbf{v}}^{(k)}$ with $k=1\ldots N$, are the
Jordan generalized eigenvectors, and satisfy
\begin{equation} \label{eq:GeneralizedEigenvectorDefinition}
  (\mathbf{L}_{\rm{cent}}-\lambda\mathbf{I})
  {\mathbf{v}}^{(k)}=
  \mathbf{L}_{\rm{cent}}
  {\mathbf{v}}^{(k)}=
{\mathbf{v}}^{(k-1)}
\end{equation}
for $k=2\ldots N$, along with the equation for the true eigenvector
$\mathbf{L}_{\rm{cent}}
  {\mathbf{v}}^{(1)}={\mathbf 0}
$. The  matrices for  $\mathbf{L}_{\rm{cent}}$ and  
 $ {\mathbf{v}}^{(k)}$
in this basis have the forms
\begin{equation}
  \label{eq:Lcentij}
  \left(L_{\rm cent}\right)_{ij}=\delta_{i,j-1}
\quad\quad
(v^{(k)})_j=\delta_{jk}
\,.   
\end{equation}
From Eq.~(\ref{eq:GeneralizedEigenvectorDefinition}) it is clear that 
the application $N$ times of $\mathbf{L}_{\rm{cent}}$ gives zero for 
any of the basis vectors, and hence for any vector. This 
again shows that the operator is nilpotent of index $N$.

The spectrum of $\mathbf{L}_{\rm{cent}}$ admits a beautifully simple
physical interpretation.  The first generalized eigenvector, the true
eigenvector, corresponds to $\delta h\neq0$ only in the outermost
shell. In an iteration cycle, we first solve for the potential inside
this outermost shell and find that the only change is that the
potential is uniformly changed by a constant, $\delta \Phi$. Since the
central density, and hence the central enthalpy, is kept fixed, we
adjust $\kappa$ in Eq.~(\ref{eq:gendeltareln}) so that $\delta h=0$ at
the origin. But $\delta \Phi$ is the same everywhere in the stellar
interior, so by setting $\delta h$ to zero at the origin, we set it to
zero everywhere. Thus an initial perturbation only in the outermost
shell is made to vanish in one cycle of iteration. This is the
physical picture of the mathematical fact $\mathbf{L}_{\rm{cent}}
{\mathbf{v}}^{(1)}={\mathbf 0} $.

The second, generalized eigenvector (${\mathbf v}^{(2)}$ in the
notation of Eq.~(\ref{eq:GeneralizedEigenvectorDefinition})), consists
of only the two outermost grid zones having $\delta h\neq0$. This
means that all zones interior to the outermost zone will have the same
change $\delta \Phi$, while the outermost zone will have a different
value of $\delta \Phi$. By a minor variation of the previous argument we
can see that one cycle of iteration will eliminate the density
perturbation except in the outermost shell, in other words, will
convert ${\mathbf v}^{(2)}$ to ${\mathbf v}^{(1)}$. The extension of  this
viewpoint
explains  all the {generalized} eigenvectors, with
${\mathbf v}^{(3)}$ having $\delta h\neq0$ only in the outermost
three zones, and so forth. (Note: For all the generalized eigenvectors
except the true eigenvector, we could set the density in the outer shell
to zero; since density only in that shell is the zero eigenvector, changing
it doesn't affect the action of the updating operator.)

\subsection{Fixed central density and the continuum}\label{fixedcentralcont}

The motivation for this paper is primarily the convergence of
numerical methods, so the considerations above for finite difference
computations suffice in practice for fixed central density iteration.
As a matter of principle, however, it is interesting to consider the
continuum equivalent of the finite difference problem of the previous
subsection.

We start by rewriting Eq.~(\ref{eq:fixcentral1}) in a notation for finding
the $p+1$ iterant from the $p$th,
\begin{equation}\label{eq:fixcentralmod}
  \delta h^{(p+1)}(r)
  =4\pi G\int_0^R \left[{\cal P}(r')\left(\frac{1}{\max(r,r')}
    -\frac{1}{r'}
  \right)
  r'^2\right]\delta h(r')^{(p)}\,dr'
  =
  -4\pi G\int_0^r \left[{\cal P}(r')r'\left(1-\frac{r'}{r}\right)\right] 
\delta h(r')^{(p)}\, dr'
  \,.
\end{equation}
The factor in square brackets in the integral is nonnegative. We let 
$\delta h^{(0)}_{\rm max}$ be the 
maximum on the interval $(0,R)$, of the initial deviation 
$|\delta h(r')^{(0)}|$ from the solution
and
${\cal P}_{\rm max}$ the maximum of ${\cal P}(r)$
on
$(0,R)$. We then have 
\begin{equation}
 | \delta h^{(1)}(r)|
  \leqslant
  \left|4\pi G\delta h^{(0)}_{\rm max}{\cal P}_{\rm max}\right|\int_0^r 
r'\left(1-\frac{r'}{r}\right)\, dr' = 
\left|4\pi G{\cal P}_{\rm max}\right|\frac{r^2}{3!}\,\delta 
h^{(0)}_{\rm max}
    \,.
\end{equation}
This inequality and Eq.~(\ref{eq:fixcentralmod}) then gives us
\begin{equation}
  |\delta h^{(2)}(r)|
  \leqslant \left|4\pi G{\cal P}_{\rm max}\right|^2\,\frac{r^4}{5!}
   \,\delta  h^{(0)}_{\rm max}\,,
\end{equation}
and
\begin{equation}\label{eq:continuumconverge}
  | \delta h^{(p)}(r)|
  \leqslant \left|4\pi G{\cal P}_{\rm max}\right|^p\,
\frac{r^{2p}\,\delta h^{(0)}_{\rm max}}{(2p+1)!} 
  \leqslant \left|4\pi G
{\cal P}_{\rm max} R^2\,\right|^p\,\frac{\delta h^{(0)}_{\rm max}}{(2p+1)!}\ .
\end{equation}
But $C^p/(2p+1)!\rightarrow 0$ as $p\rightarrow\infty$ for any finite
$C$, hence the iteration defined by Eq.~(\ref{eq:fixcentralmod})
converges.  Unlike the discrete case, the continuum operator is not
nilpotent, since the results of operating on a function a finite
number of times does not give zero.  The name ``quasinilpotent,''
however, is sometimes applied to an operator like $L_{\rm cent}$ for
which the spectrum consists only of zero.

We can also inquire about the eigenvector problem for the continuum
\begin{equation}\label{eq:conteigen}
  -4\pi G\int_0^r \left[{\cal P}(r')r'\left(1-\frac{r'}{r}\right)\right]  
v(r')\, dr'\equiv L_{\rm cent}(v)=\lambda v\,.
\end{equation}
If we assume that $v(r)$ is bounded, and $\lambda\neq0$ then the same
argument used to arrive at Eq.~(\ref{eq:continuumconverge}) tells us
that when $L_{\rm cent}$ is applied $n$ times to $v$ we get
\begin{equation}
  |v|< \left|\frac{4\pi G {\cal P}_{\rm max} R^2}{\lambda}\right|^p\,
\frac{v_{\rm max} }{(2p+1)!}
  \,,
\end{equation}
which vanishes as $p\rightarrow\infty$ showing that no bounded
eigenfunction with $\lambda\neq0$ can exist.

We next show that no bounded eigenfunction with $\lambda=0$ can
exist. Since ${\cal P}(r')r'(1-r'/r)$ is nonnegative in the integrand of
Eq.~(\ref{eq:conteigen}), $v(r)$ must change sign in the integral.
Let us assume that, after $r=0$, the eigenfunction $v(r)$ has its
smallest zero at $r=a$. For definitiveness we take $v$ to be positive
on $(0,a)$.  From Eq.~(\ref{eq:conteigen}), we have
\begin{equation}
  \label{eq:firstzero}
\lambda v(a)= - 4\pi G\left[\int_0^a r'\left(1-\frac{r'}{a}\right){\cal P}(r') v(r')
dr'\right]\,.
\end{equation}
In the integrand the factor ${\cal P}(r')(1-r'/a)$ is
nonnegative and not identically zero, and by hypothesis $v$ is
nonnegative and not identically zero. It follows that $v(a)$ cannot be
zero. Since this contradicts our assumption about a zero at $a$, we
conclude that $v(r)$ cannot have a zero in $(0,R)$, and hence a bounded 
eigenfunction cannot exist. 

If we relax the condition that the eigenfunction must be bounded, and
allow distributional solutions, we immediately see that the delta
function $v=\delta(r-R)$ is an eigensolution since it satisfies
\begin{equation}
  \label{eq:deltafn}
  L_{\rm cent}\big(\delta(r-R)\big)=
4\pi G\left[\int_0^r r'\left(\frac{r'}{r}-1\right){\cal P}(r')\,\delta(r'-R)\,
dr'\right]=0.
\end{equation}
This eigensolution, of course, is the continuum equivalent of the ``outer
shell only'' eigenvector of the discrete problem.

It is interesting to consider the analog in the continuum of
the Jordan canonical form\cite{arnold}. This would require a
definition of the functions that constitute our Banach space on which
the operator $L_{\rm cent}$ operates. Without going into such detail
we can make some interesting observations about a Jordan-like 
decomposition for $L_{\rm cent}$. To start we note that in an $N$
dimensional context the Jordan basis (for a single zero eigenvalue
Jordan block) can be constructed starting with the eigenvector
$v^{(1)}$ and proceeding with an inverse $L^{-1}_{\rm cent}$ operator. 
(This inverse, of course, is not unique, but we can choose it always to 
give a result orthogonal to $v^{(1)}$.) With this inverse we
construct
\begin{equation}
  v^{(2)}=L^{-1}_{\rm cent}(v^{(1)})\quad\quad
  v^{(3)}=L^{-1}_{\rm cent}(v^{(2)})\quad\quad\cdots\quad\quad
  v^{(N)}=L^{-1}_{\rm cent}(v^{(N-1)}).
\end{equation}

If we attempt to follow this pattern in the continuum
we can use the 
inverse of $L_{\rm cent}$ to be 
\begin{equation}\label{Lcentinverse}
  L_{\rm cent}^{-1}(f)=-\,\frac{1}{4\pi G {\cal P}(r) r^2}\frac{d}{dr}\left(r^2\frac{d}{dr}f\right)\,.
\end{equation}
With $v^{(1)}=\delta(r-R)$, the analogous sequence of distributions is given by 
\begin{equation}
  v^{(2)}=L_{\rm cent}^{-1}(v^{(1)})
  \quad\quad
 v^{(3)}=L_{\rm cent}^{-1}(v^{(2)})
\quad\cdots\quad
 v^{(k+1)}=L_{\rm cent}^{-1}(v^{(k)})\,.
\end{equation}
Such a sequence -- technical objections aside -- would give a basis
with the property in Eq.~(\ref{eq:GeneralizedEigenvectorDefinition}).
The technical objection, of course, is that the eigenfunction is a
delta function, so that our sequence would consist of more and more
singular generalized functions. Worse, this sequence in no way
resembles the finite dimensional Jordan basis in which each subsequent
basis vector is an outer shell that is ``thicker'' than the previous
basis vector.

A more interesting sequence consists of the functions 
\begin{equation}\label{eq:procedure}
  v^{(0)}=1\quad\quad v^{(-1)}=L_{\rm cent}(v^{(0)})  \quad\quad 
v^{(-2)}=L_{\rm cent}(v^{(-1)}) 
  \quad\quad v^{(-3)}=L_{\rm cent}(v^{(-2)}) \quad \cdots \,.
\end{equation}
This sequence formally satisfies the Jordan basis criterion in
Eq.~(\ref{eq:GeneralizedEigenvectorDefinition}) for
$k=0,-1,-2,\cdots\,.$ The limit of this sequence, ``$v^{(-\infty)}$''
should in some sense represent the single eigenfunction. That is,
$v^{(-k)}$ should approach the delta function at the stellar surface
as $k\rightarrow\infty$.

We let a simple example suffice to show that, in a rough sense, this is the case.  For
the $n=1$ polytropic equation of state we have from
Eqs.~(\ref{eq:polyEOS2}) and (\ref{Hdef}) that ${\cal P}=1/2K$, so that
the procedure of Eq.~(\ref{eq:procedure}) gives
\begin{equation}
    v^{(0)}=1\quad\quad    v^{(-1)}=\left(\frac{-2\pi G}{K}\right)\,\frac{r^2}{3!}\quad\quad
v^{(-2)}=\left(\frac{-2\pi G}{K}\right)^2\,\frac{r^4}{5!}\quad\quad
v^{(-p)}=\left(\frac{-2\pi G}{K}\right)^{p}\,\frac{r^{2p}}{(2p+1)!}\,.
\end{equation}
In intuitive accord with the finite dimensional case, and with the physical picture,
as $p\rightarrow\infty$ the generalized eigenfunction, in a rough sense, approaches
a density profile that is concentrated at the outer boundary.

\subsection{Fixed intermediate density and finite difference computations }

We now consider the case of spherical models with density fixed at
distance $r_f\leqslant R$ from the center. 
With $\delta h|_{r_f}=0$ in 
Eqs.~(\ref{eq:gendeltareln}), (\ref{eq:gendeltaU}) we arrive at
\begin{equation}  \label{eq:fixrzero}
\delta h^{\rm new}(r)
=4\pi G\left[\int_0^R {\cal P}(r')\frac{\delta h(r')^{\rm old}}{\max(r,r')}\,
r'^2\,dr'-\int_0^R {\cal P}(r')\frac{\delta h(r')^{\rm old}}{\max(r_{f},r')}\,
r'^2\,dr'\right]
\equiv L_{r_{f}}(\delta h^{\rm old})
\,.
\end{equation}
We discretize as in Eq.~(\ref{rdiscrete}), and for convenience we
choose $r_{f}=(f-\tfrac{1}{2})\,\Delta r$ where $f$ is a positive
 integer $\leq N$.  In place of
Eq.~(\ref{eq:LcentDiscreteDefinition}) we now have
\begin{equation} \label{eq:LintDiscreteDefinition}
  (L_{r_f})_{ij}=4\pi G {\cal P}(r_j)\, \left(\frac{1\,}{{\rm max}(r_i,r_{j})}\,
-\frac{1\,}{{\rm max}(r_f,r_{j})}\,
\right)r_{j}^2\Delta r\,. 
\end{equation} 
We notice that this matrix has the structure
\begin{equation}\label{Lro}
  \mathbf{L}_{r_f}=\left[
    \begin{array}{c|c}
      \mathbf{S}\mathbf{D}&\mathbf{0}\\ \hline
\mathbf{M}&\mathbf{T}
    \end{array}
\right]\,,
\end{equation}
where: (i) $\mathbf{S}\mathbf{D}$ is a  $(f-1)\times (f-1)$ square matrix consisting of a
symmetric matrix right multiplied by a diagonal matrix; (ii) $\mathbf{T}$ is a
strictly lower triangular $(N-f+1)\times(N-f+1)$ square matrix; (iii)
$\mathbf{M}$ is a $ (N-f+1)\times(f-1)$ matrix.
To discuss eigensolutions we write column eigenvectors in the form
\begin{equation}
\mathbf{v}=\left[
  \begin{array}{c}
    \mathbf{u}\\ \mathbf{w}
  \end{array}
\right]\,,
\end{equation}
where $\mathbf{u}$ is a column of length $f-1$ and $\mathbf{w}$ is a column of length
$N-f+1$, 
so that
\begin{equation}
  \left[
    \begin{array}{c|c}
      \mathbf{S}\mathbf{D}&\mathbf{0}\\ \hline
\mathbf{M}&\mathbf{T}
    \end{array}
\right]\,\left[
  \begin{array}{c}
    \mathbf{u}\\ \mathbf{w}
  \end{array}
\right]\,=
\left[
  \begin{array}{c}
    \mathbf{S}\mathbf{D}\;\mathbf{u}\\ \mathbf{M}\;\mathbf{u}+\mathbf{T}\;\mathbf{w}
  \end{array}
\right]
\,.
\end{equation}

For vectors with $\mathbf{u}=\mathbf{0}$ the eigenproblem reduces to
\begin{equation}
  \mathbf{T}\;\mathbf{w}=\lambda \mathbf{w}\,.
\end{equation}
Since $\mathbf{T}$ is strictly lower triangular we have, from the discussion 
in Sec.~\ref{subsecthreeA}, that the only eigenvalue is zero, and that 
there is only a single eigenvector. The rest of the $N-f+1$ dimensional 
space on which $\mathbf{T}$ operates is spanned by generalized eigenvectors, as 
in Sec.~\ref{subsecthreeA}.

We next consider solutions of the
$(f-1)\times (f-1)$ problem 
\begin{equation}
  \mathbf{S}\mathbf{D}\ \mathbf{u}=\lambda \mathbf{u}\,.
\end{equation}
We have seen in Sec.~\ref{sec:fixkappaFD} that the eigenvectors for
this problem are complete in the $(f-1)\times(f-1)$ subspace, hence
there exist $f-1$ eigenvectors in the $(f-1)\times(f-1)$ sector with,
in general, distinct eigenvalues.  Let $\mathbf{u}^{(k)}$, with $k=1,2,\cdots
f-1$ represent the set of these column vectors of length $f-1$, and
let $\lambda_{k}$ represent the corresponding eigenvalues.

The next step is to define $\mathbf{w}^{(k)}$ to be the solution of 
the $(N-f+1)\times(N-f+1)$ matrix equation
\begin{equation}\label{phieq}
\left(  \mathbf{T}-\lambda_{k}\mathbf{I}\right)\mathbf{w}^{(k)}=-M\mathbf{w}^{(k)} \quad\quad k=1,2\cdots f-1\,,
\end{equation}
with $\mathbf{I}$ the unit matrix.  If we assume that $\mathbf{S}\mathbf{D}$ has no zero
eigenvectors (which is true for all models numerically checked) then
$\mathbf{T}-\lambda_{k}\mathbf{I}$ is invertible, since the only eigenvalue of $\mathbf{T}$ is zero.
This guarantees that  solutions of Eq.~(\ref{phieq}) exist for all 
$(N-f+1)$ values of $k$. The column vectors combining $\mathbf{u}^{(k)}$ and $\mathbf{w}^{(k)}$
are then $(N-f+1)$ eigenvectors, since
\begin{equation}\label{km1eigens}
  \left[
    \begin{array}{c|c}
      \mathbf{S}\mathbf{D}&\mathbf{0}\\ \hline
\mathbf{M}&\mathbf{T}
    \end{array}
\right]\left[
  \begin{array}{c}
    \mathbf{u}^{(k)}\\ \mathbf{w}^{(k)}
  \end{array}
\right]
=
\left[
  \begin{array}{c}
    \mathbf{S D}\;\mathbf{u}^{(k)}\\ \mathbf{M}\;\mathbf{u}^{(k)}+\mathbf{T}\;\mathbf{w}^{(k)}
  \end{array}
\right]
=
\left[
  \begin{array}{c}
    \lambda_{k} \mathbf{u}^{(k)}\\ \lambda_{k}\mathbf{w}^{(k)}
  \end{array}
\right]
\,.
\end{equation}
These $f-1$ eigenvectors, along with the single zero-eigenvalue
eigenvector, are the total set of eigenvectors of $\mathbf{L}_{r_f}$. It is
clear from previous discussions that convergence of the iteration will
depend on whether any of the nonzero eigenvalues has a magnitude
greater than unity.

\subsection{Fixed intermediate density and  continuum models} 

For additional insight into the the numerically relevant finite difference
models of 
 the previous
section we now consider the continuum perturbation problem with
\begin{equation}
  \delta\kappa=-\left.\delta \Phi\right|_{r_{f}}\ .
\end{equation}
It is convenient to view this in terms of the 
differential operator that is the inverse of the updating operator. 
As in Eq.~(\ref{eq:inverseL}), we have
\begin{equation}
  \label{eq:inverseLrzero}
L_{r_{f}}^{-1}\big(v\big) = -\frac{1}{4\pi Gr^2 {\cal P}(r)}\,\frac{d}{dr}\left(
r^2\frac{dv}{dr}
\right)\,,
\end{equation}
and the eigenequation is
\begin{equation} \label{eq:rzeroSturmLiouville}
  L_{r_{f}}^{-1}\big(v\big)=(1/\lambda)v\,.
\end{equation}
The conditions on $v(r)$ at $r\rightarrow0$ are as before, but now 
the other condition on the eigensolution is that $v(r_f)=0$.
For the inner product
\begin{equation}
  \label{eq:rzerodotprod}
    v_1\cdot v_2=\int_0^{r_f} r^2{\cal P}(r)v_1(r) v_2(r)\,dr\,,
\end{equation}
this constitutes a Sturm-Liouville problem, and hence 
$L_{r_{f}}^{-1}
$
has a complete
eigenbasis. More specifically, it has an eigenbasis that is complete
in the mean (with respect to the above inner product) on the interval
$(0,r_{f})$. But the interval relevant to the stellar interior is
$(0,R)$, and there is no reason that the eigensolutions will be
complete in any meaningful sense on $(0,R)$.

The continuum eigenvectors for the interval $(0,r_{f})$ correspond to
the eigenvectors $\mathbf{u}^{(k)}$ of the $(f-1)$ dimensional subspace in the
discrete implementation of the problem.  Just as the solutions to
Eq.~(\ref{eq:inverseLrzero}) have no special significance for
$r_{f}<r< R$, the column eigenvectors $\mathbf{v}^{(k)}$ of the discrete
problem, in Eq.~(\ref{km1eigens}), have no special significance for
the bottom $N-f+1$ elements of the column.

For $r_{f}< r< R$, the continuum problem has
similarities to the $r_{f}=0$ fixed central density problem discussed in
Secs.~\ref{subsecthreeA} and \ref{fixedcentralcont}. In particular,
$\delta(r-R)$ is an eigenfunction (or generalized function) with zero
eigenvalue, and -- as in Secs.~\ref{subsecthreeA} and
\ref{fixedcentralcont} -- the generalized eigenvectors of the Jordan
decomposition have analogs in the continuum.
The physical picture of the zero eigenvector
applies just as in
Secs.~\ref{subsecthreeA} and \ref{fixedcentralcont}.

By specializing to the $n=1$ case, we may once again 
benefit from a simple closed-form example.
With the notation of Eq.~\eqref{eq:dimnless},
the operator \eqref{eq:fixrzero}  becomes
\begin{eqnarray} \label{eq:Lrzero1} \nonumber
L_{r_{f}}(v)\,&\equiv& \ \int_0^{\pi}\ v(\xi')
\left(\frac{1\,}{\max(\xi,\xi')}\,-\frac{1\,}{\max(\xi_{f},\xi')}\right)
\,\xi '^2\,d\xi' 
\\
&=& \int_0^{\xi_{f}}\ v(\xi')
\left(\frac{1\,}{\max(\xi,\xi')}\,-\frac{1\,}{\xi_{f}}\right)
\,\xi '^2\,d\xi'
+\int_{\xi_{f}}^{\pi}\ v(\xi')
\left(\frac{1\,}{\max(\xi,\xi')}\,-\frac{1\,}{\xi'}\right)
\,\xi '^2\,d\xi' \,.
\end{eqnarray}
For this operator, $\delta(\xi-\pi)$ is an eigenvector (in the vector
space of distribution functions on $\mathbb{R}$) with zero eigenvalue,
and the solutions $v^{(k)}$ to
\begin{equation}
  L_{r_{f}}(v^{(k)})=\lambda_{k}\; v^{(k)},
\end{equation}
given by 
\begin{equation}
  v^{(k)}(\xi)=
  \xi^{-1} \sin({\lambda_{k}^{-1/2}\xi}),\quad\quad  
 \lambda_{k}^{-1/2}=k\pi/\xi_f,\quad k=1,2,3\cdots\,,
\end{equation}
are eigenvectors. This suggests that, for an $n=1$ polytrope, the
iteration \eqref{eq:fixrzero} should converge for any choice of
$\xi_f<\pi$, but the convergence rate is maximized when the density is
fixed at the center. We note, however, that the above eigenvectors are
{\em not} complete on $(0,\pi)$ and that generalized eigenvectors can
be constructed using the procedure of Eq.~(\ref{eq:procedure}).
\section{Summary and Conclusions} 
\label{sec:summary}

We have investigated the properties of the updating operator for
linearized iteration of the two equations that govern Newtonian
neutron star structure, and have focused on spherically symmetric
models. We have considered two constraints on the iteration: (i) the
injection energy $\kappa$ is held fixed, and (ii) density is held
fixed at a specified radius.

In the case of fixed-$\kappa$ iteration we have found that both the
finite dimensional problem (for finite difference discretization) and
the continuum problem are self-adjoint and convergence is determined
by the spectrum of its eigenvalues. For polytropic equations of state,
numerical work had always led to divergence of iteration. We have
shown, in fact, that there is a rigorous bound on the largest
eigenvalue of the updating operator: it must be greater than the
polytropic index $n$. Since all numerical experiments had been carried
out with $n>1$, this meant that there had to be an updating eigenvalue
greater than unity, and hence that iteration would diverge. This did
suggest, however, that for polytropic equations of state with
unphysically small values of $n$, convergence might be
possible. Numerical experiments showed that in fact this is the case,
thereby verifying the applicability of the analysis.

The updating operator for fixed central density was shown to have a
very different nature than that for fixed-$\kappa$. In the finite
dimensional case corresponding to a finite difference representation
of the equations, there is only a single eigenvector, with eigenvalue
zero, and the updating operator is nilpotent. The continuum version of
the fixed central density problem is not connected as directly to the
finite difference problem as in the fixed-$\kappa$ case. We have
shown, however, that for the continuum iteration converges. It is also
possible to construct a sequence of functions that have some of the
spirit of the generalized eigenbasis of the Jordan decomposition of
the finite dimensional problem.  

For density fixed at some radius $r_{f}$ other than the center, the
updating operator, not surprisingly, has mixed properties. The space
on which the updating operator acts can be separated into two sectors,
one corresponding to radius $\leqslant r_{f}$ on which the updating
operator acts more-or-less like the updating operator in the
fixed-$\kappa$ case; and the other sector, for radius $>r_{f}$
on which the updating operator acts more-or-less like the operator for
fixed central density.

Although only the finite difference analyses are directly applicable
to numerical iteration, the connection to the continuum is not only
useful, but has a practical importance: it suggests that the
properties of the updating operator are not idiosyncrasies of finite
differences. It thus adds confidence that, for example, the use of a
Gaussian method for the integral will converge or diverge just as the
finite difference case would.

The analyses presented here have been limited to spherical
symmetry. But further work, mostly numerical, will be reported
elsewhere that shows that many of the general conclusions reported
here also apply to rotating stars and to binaries. In particular, the
fixed-$\kappa$ updating operator is always self-adjoint, and generally
has an eigenvalue greater than unity; and rotating stars with fixed
rotation speed and fixed central density lead to a nilpotent updating
operator.

Although the main motivation for the work undertaken here has been a
mathematical understanding of iteration properties, the results have
potentially useful applications. In particular, the convergence of a
nilpotent updating operator (like the fixed central density operator)
is very different from that of a self-adjoint updating operator (like
the fixed-$\kappa$ operator for a polytropic equation of state with a
small polytropic index). The nilpotent operator will reach a solution
at machine precision within a finite number of iterations, while
convergent iteration in general may approach the correct solution
gradually, and slowly. It may also be useful to understand the nature
of the iteration when a test must be made whether or not the process
is converging. For convergent iteration with a self-adjoint updating
operator a simple measure of the difference in subsequent solutions
can be used, employing the same metric for which the operator is
self-adjoint.  With an updating operator like that for fixed central
density, convergence may give an early appearance of divergence. If
the initial perturbation is a distribution concentrated near the outer
edge of the stellar model, each iteration will move the perturbation
inward, reducing it only after many steps. Too simple a test for
convergence might misinterpret this iteration as nonconvergent.


\section{Acknowledgments}
We gratefully acknowledge support for this work under NSF grants
PHY-0554367, PHY-0503366, NASA grant NNG05GB99G, by the Greek State
Scholarships Foundation, and by the Center for Gravitational Wave
Astronomy. We thank Alan Farrell for supplying some computational
results.


\bibliographystyle{prsty}


\end{document}